# A unified model of density limit in fusion plasmas

P. Zanca[1], F. Sattin[1], D. F. Escande[2], G. Pucella[3], O. Tudisco[3]

[1]*Consorzio RFX (CNR, ENEA, INFN, Università di Padova, Acciaierie Venete Spa) Corso Stati Unit 4, 35127 Padova (Italy)*

[2]*Aix-Marseille Univ, CNRS, PIIM, Marseille, France*

[3]*C.R. ENEA Frascati, CP65, 00044 Frascati, Italy*

**Abstract**

A limit for the edge density, ruled by radiation losses from light impurities, is established by a minimal cylindrical magneto-thermal equilibrium model. For ohmic tokamak and reversed field pinch the limit scales linearly with the plasma current, as the empirical Greenwald limit. The auxiliary heating adds a further dependence, scaling with the 0.4 power, in agreement with L-mode tokamak experiments. For a pure externally heated configuration the limit takes on a Sudo-like form, depending mainly on the input power, and is compatible with recent Stellarator scalings.

A discharge-terminating density limit (DL) is found in all the magnetic confinement fusion devices [1]. One of the main interpretative branches invokes impurity radiation losses, which scale with the square of the density. The consequent cooling of the plasma can become critical at high density, giving rise to a variety of instabilities, both thermal [2-8], and MHD [9-13]. Given the rich phenomenology, DL seems elusive of an explanation based on a single mechanism. This letter presents a complementary approach to the problem, analysing, in cylindrical geometry, the feasibility of a magneto-thermal equilibrium with realistic temperature profile, rather than addressing specific instabilities. Such a study provides a unified interpretation of the phenomenon, given that a DL ruled by light impurities radiation (experiments show that any significant contamination by heavy impurities is just detrimental towards the achievement of high densities [1]), quantitatively consistent with experimental scalings, emerges naturally for all the magnetic configurations. In particular, we found a Greenwald-like scaling [1] for tokamak and reversed field pinch (RFP), and a Sudo-like scaling [14, 15] for a pure externally heated configuration, taken as approximation of the stellarator. We are aware that this analysis cannot exhaust the topic, since some instability mechanism is necessary to describe the dynamical route to the plasma termination. Consequently, we speak of an 'equilibrium DL' and not of the DL in the ultimate sense. This work has been inspired by analyses of the ohmic tokamak presented in [16] (section 7.8) and in [17] (section 8). The differences rely in a more general approach, besides a more accurate treatment of the profile dependent terms.

We introduce a *minimal cylindrical equilibrium model* (each quantity depending on the radial coordinate *r* only), analytically treated with a formalism able to unify the configurations with an applied electric field, namely the tokamak, both ohmic and with additional heating, and the RFP, considered as purely ohmic. Basically, we will take integral relations from the heat transport equation, in some way similar to those carried out in [18] apart for the simpler geometry, and combine them with Ohm's law at *r*=0 (on-axis). Ohm's law is replaced by a suitable scaling for the energy confinement time in the case of pure auxiliary heating.

Thermal balance is described by the sum of the heat transport equations for electrons and main ions under stationary condition [19]:

$$1) \quad \frac{d}{dr}\left(rK\frac{dT}{dr}\right) + r[\wp_{Ohm} + \wp_{aux} - n_e n_I Rad(T)] = 0, \quad \wp_{Ohm} = \eta J_\Omega^2$$

Here, $T$ is the electron temperature and $K$ an effective conductivity, which in terms of ions and electrons densities ($n_i$, $n_e$), temperatures ($T_i$, $T$) and diffusivities ($\chi_i$, $\chi_e$) can be written as $K = 1.6 \times 10^{-16} \times n_e \chi_e [1 + n_i \chi_i T_i'/(n_e \chi_e T')]$, $' \equiv d/dr$. Throughout this paper we use the International System (SI) of units, except for the temperature expressed in keV. Therefore, $K$ incorporates the numerical factor $1.6 \times 10^{-16}$. The terms $\wp_{Ohm}$, $\wp_{aux}$ are heating power densities, ohmic and auxiliary respectively. The total current **J** may include the contribution **J**$_\mathbf{d}$ driven by the auxiliary heating, in addition to the ohmic component **J**$_\mathbf{\Omega}$. Low-β condition is assumed, so the currents are approximated parallel to the magnetic field. Accordingly, $\eta$ is the parallel resistivity. Moreover, **J** and **J**$_\mathbf{\Omega}$ are taken in the same direction, so that the absolute values are related by $J=J_\Omega \pm J_d$. The cooling rate *Rad(T)* determines the radiation emission and depends on the atomic physics of the impurities, whose total density is denoted by $n_I$. Here *Rad(T)* represents a weighted average between the different species involved. For light impurities, it is strongly peaked at low temperatures (order tens eV), with a secondary, much smaller maximum at higher temperatures [20]. Accordingly, a step-wise model is here adopted: $Rad(T \leq T_*)=R_*$, $Rad(T>T_*)=0$, being $T_*$ of the order of 30eV. By taking a monotonic decreasing temperature profile, satisfying $T'(0)=0$ on-axis, and $T(a)=0$ at the wall radius, $Rad(T)$ becomes a function of $r$: $Rad(r \geq r_*)=R_*$, $Rad(r<r_*)=0$, where $T(r_*)=T_*$. Integration of (1) is then straightforward. In particular, the half-sum of the integrals of (1) in $[0, a]$ and $[0, r_*]$ provides a balance between heat-flux, power and radiation losses:

2) $\quad K_* X + Q(a) - \dfrac{\Re_*}{2\,a} = 0$

Here $\quad X = [T'(a) + T'(r_*)]/2 \quad , \quad Q(r) = r^{-1}\displaystyle\int_0^r (\wp_{Ohm} + \wp_{aux})\rho\,d\rho \quad ,$

$\Re_* = R_* \displaystyle\int_{r^*}^a n_e\, n_I\, \rho\,d\rho \cong R_* f_* \displaystyle\int_{r^*}^a n_e^2 \rho\,d\rho$, and the constants $K_*$, $f_*$ approximate $K$ and the impurity concentration $f_I = n_I/n_e$ within the radiative layer $[r_*, a]$. Given the smallness of $T_*$, we use repeatedly the approximation $r_* \approx a$. Accordingly, $Q(a) \approx Q(r_*)$. Moreover, $Q(a) = a \langle \wp \rangle / 2$, being $\langle \wp \rangle$ the volume-average heating power density. Nonetheless, it is correct to distinguish the temperature derivatives at $a$ and $r_*$ within $X$, since a finite variation of $T'$ occurs in the radiative layer: $T'(a) - T'(r_*) \approx \Re_* / (a\,K_*)$. Therefore, a Taylor expansion of $T(r)$ about $r=a$ must retain also the second-order derivative term:

3) $\quad T_* \cong T'(a)(r_* - a) + T''(a)(r_* - a)^2 / 2 \cong X(r_* - a)$

Note that $X$ must be negative in order for $T_*$ to be positive. Defining a typical density $n^*$ of the emission layer by $(n^*)^2 a(a - r_*) = \displaystyle\int_{r^*}^a n_e^2 \rho\,d\rho$ and making use of (3), condition (2) becomes a second-order algebraic equation for $X$:

4) $K_* X^2 + \frac{a}{2} \langle \wp \rangle X + \frac{1}{2} R_* T_* f_* (n^*)^2 = 0$.

The discriminant is positive or zero, and there are real-valued negative solutions for X, when

5) $n^* \leq a \langle \wp \rangle (8 K_* R_* T_* f_*)^{-1/2}$

Equation (5) is already suggestive of a DL relation, but it will be further expanded into more explicit forms, applying it to the different configurations. One can check that $T'(a) = 0$ when (5) is satisfied as equality: in this somewhat idealized model, DL corresponds to the condition of vanishing heat flux at the edge, implying radiation emission equal to the input power.

*Tokamak and RFP*

Three further equations are needed in these cases. The first is on-axis Ohm's law, here written in a general, compact form:

6) $E_\phi = \eta(0) J_{\Omega,\phi}(0) \times C(0)^{h_{RFP}} = \eta_1 Z_{eff}(0) \, T(0)^{-3/2} \frac{B_\phi(0)}{2\pi R_0 \, q(0)} \times C(0)^{h_{RFP}} \times \xi(0)^{h_{AUX}}$,

$\xi = J_\Omega / J$, $\quad \eta_1 = 1.65 \times 10^{-2} \ln \Lambda$

This expression is now explained. In stationary conditions $E_\theta = 0$ and $E_\phi(r) = const = E_\phi$. Resistivity is expressed by the classical Spitzer law (neo-classical effects vanish at $r=0$) [16, 21]: $Z_{eff}$ is the effective ion charge and $ln\Lambda$ the Coulomb logarithm. The total current satisfies $\mu_0 \mathbf{J} = \sigma(r) \mathbf{B}$, with $\sigma(0) = 2/(q(0) R_0)$ and $q = r B_\phi / (R_0 B_\theta)$. The two exponents $h_{RFP}$ and $h_{AUX}$ are configuration selectors. For ohmic tokamak, $h_{RFP}=0$, $h_{AUX}=0$. For tokamak with auxiliary heating, $h_{RFP}=0$, $h_{AUX}=1$: the function $\xi(r)$ relates the absolute values of ohmic and total current (at $r=0$, $J=J_\phi$, $J_\Omega=J_{\Omega,\phi}$). For the RFP, $h_{RFP}=1$, $h_{AUX}=0$: the function $C(r)$ (anomaly function), defined by $\mathbf{E} \cdot \mathbf{B} = C \eta \mathbf{J} \cdot \mathbf{B}$, or equivalently by $\wp_{Ohm} = E_\phi J_\phi / C$, encapsulates the 1D effect of the dynamo mechanism, which acts by 2D or 3D MHD perturbations. Numerical simulations estimate $C$ larger than 1 in the plasma core, decreasing below this value only at the edge [22, 23].

The second relation needed is a compact expression for the heating power density in equation (1):

$$7)\quad \wp_{Ohm} + \wp_{aux} = \eta J_\Omega^2 + \wp_{aux} \equiv \frac{E_\phi J_\phi}{C^{h_{RFP}} \times \Gamma^{h_{AUX}}}$$

A further profile function $\Gamma(r)$ encapsulates the additional heating contribution. Note that, definition (7) recovers the correct expressions for ohmic tokamak and RFP, respectively $E_\phi J_\phi$ and $E_\phi J_\phi/C$. Formulation (7) is convenient, because it allows exploiting the constancy of $E_\phi$ and Ampère's law $\mu_0 J_\varphi = r^{-1} d(rB_\theta)/dr$ to write

$$8)\quad Q(r) = \frac{B_\theta(a)}{\mu_0} \frac{E_\phi}{C(0)^{h_{RFP}} \Gamma(0)^{h_{AUX}}} \psi(r)$$

Here, $\psi(r) = [r\hat{B}_\theta(a)]^{-1} \int_0^r d(\rho \hat{B}_\theta)/d\rho \; \hat{C}^{-h_{RFP}} \hat{\Gamma}^{-h_{AUX}} d\rho$, with $\hat{B}_\theta = B_\theta/B_\phi(0)$, $\hat{C} = C/C(0)$, $\hat{\Gamma} = \Gamma/\Gamma(0)$. Using (8) to express $\langle \wp \rangle = 2/a \; Q(a)$, DL (5) becomes

$$9)\quad n^*/n_G \leq 10^{-14} \times a \frac{E_\phi}{C(0)^{h_{RFP}} \Gamma(0)^{h_{AUX}}} \psi(a) \left(8K_* R_* T_* f_*\right)^{-1/2}$$

This formalism introduces the Greenwald density $n_G = 10^{14} \times I_p/(\pi a^2)$ [1], being $I_p$ the plasma current, in a natural way. The third needed equation is a further integral relation derived from (1), this time integrated twice, first over $[0,r]$, and then over $[0,a]$, after a division by $r$. In order to model profile effects in $K$, we define $\hat{K} = K/K_*$. This provides an approximate relation for the on-axis temperature:

10) $\quad K_* T(0) \approx \dfrac{B_\phi(0)}{\mu_0} \dfrac{E_\phi}{C(0)^{h_{RFP}} \Gamma(0)^{h_{AUX}}} \dfrac{a^2}{R_0} \hat{\Im}, \qquad \hat{\Im} = \dfrac{R_0}{a^2} \hat{B}_\theta(a) \int_0^a \dfrac{\psi(r)}{\hat{K}(r)} dr$

The contribution of the radiation term to (10) is smaller than $R_* f_* \int_{r^*}^{a} r^{-1} \left( \int_{r^*}^{a} n_e^2 \rho\, d\rho \right) dr = \Re_* \ln(a/r^*)$, and is discarded due the assumption that $r_*$ is close to $a$. By combining (6) and (10) to eliminate $T(0)$, a scaling law is got for $E_\phi$, containing, in particular, a factor $C(0)^{h_{RFP}} \times K_*^{3/5}$. Therefore, such a scaling cancels out the dependence on $C(0)$ and almost totally that on $K_*$ in (9):

11)
$$n^*/n_G \leq 1.54 \times 10^{-2}\, \eta_1^{2/5}\, a^{-1/5}\, R_0^{1/5}\, Z_{eff}(0)^{2/5}\, f_*^{-1/2}\, \tilde{R}^{-1/2}\, K_*^{1/10}\, q(0)^{-2/5}\, \hat{\Im}^{-3/5}\, B_\phi(0)^{-1/5}\, \psi(a)\, [\xi(0)/\Gamma(0)]^{2h_{AUX}/5}$$

We have conveniently defined $R_* T_* = \tilde{R} \times 10^{-33}$, since $\tilde{R}$ turns out to be order unity for actual cooling rates [20]. Given its weak exponent, $\eta_1$ can be safely estimated by taking just one reference value for $ln\Lambda$: $ln\Lambda=15$. The tenuous dependence upon $K_*$ in (11), not disclosed in previous derivations [16, 17], will make the final scaling very similar for tokamak and RFP, despite the different transport properties of these two devices. Now we further manage $\Gamma(0)$ and $K_*$, to make scaling (11) more transparent. First, $\Gamma(0)$ can be given a clearer form, at the expense of loosing the full generality of the formulation. Let's take the radial profiles of $\wp_{aux}$ and $J_d$ to be comparable to that of the total current $J$. This hypothesis, reasonable at least for NBI heating [24, 25], makes $\xi$ and $\Gamma$ almost radially constant. In fact, $\xi=1\pm J_d/J\approx const\approx\xi(0)$. Moreover, from definition (7) ($h_{RFP}=0$, $h_{AUX}=1$), relation $E_\phi J_\phi=\mathbf{E}\cdot\mathbf{J}=\eta J_\Omega J=\eta J_\Omega^2/\xi$, and $J\approx J_\phi$ (low-$\beta$, large aspect ratio tokamak ordering), one gets:

12) $\quad \Gamma^{-1} = \xi + \wp_{aux}/E_\phi J_\phi \approx const \approx \xi(0)\ P_{tot}/P_{Ohm}$

Here, $P_{tot} = P_{aux} + P_{Ohm}$, $P_{Ohm} = \int \eta J_\Omega^2 d^3x = \int \xi E_\phi J_\phi d^3x$, and $P_{aux}$ are respectively the total, ohmic and auxiliary global heating powers.

Furthermore, the faint dependence $K_*^{1/10}$ implies that any reasonable model of heat transport for tokamak and RFP can be adopted without affecting conclusions seriously. Let's start from the approximate relation

13) $\quad \kappa \times K_* \approx 0.5 \times 1.6 \times 10^{-16} \times a^2 \bar{n}_e / \tau_E$

where $\kappa$ is a suitable shape factor, normalized to 1 for a radially constant $K$, $\bar{n}_e$ is the line-average density, and $\tau_E = \langle 3/2\, p \rangle / \langle \wp \rangle$ is the energy confinement time (symbol $\langle \; \rangle$ denotes the volume average and $p$ the plasma pressure). Numerical solutions of the model's equations suggest the corrective pre-factor 0.5 in the r.h.s. As far as the tokamak is concerned, we take the simple neo-Alcator scaling $\tau_{E;neoAlc}(s) = 7 \times 10^{-3} \bar{n}_e (10^{-19} m^{-3}) a R_0^2 q(a)$, established at low density [16], downscaling it by a factor 2 for the high density regime we are dealing with, as suggested by the analysis presented in [26]. In this way, (12) and (13) provide the following form to (11):

14) $\quad n*_{tok} \leq 0.3 \times a^{-1/10} Z_{eff}(0)^{2/5} f_*(\%)^{-1/2} \tilde{R}^{-1/2} B_\phi(0)^{-1/5} \Psi_{tok} \left[ \xi(0)^2\, P_{tot}/P_{Ohm} \right]^{0.4 h_{AUX}} \times n_G$

The impurity concentration is expressed in percentage here. Quantity $\Psi_{tok} = q(0)^{-2/5} q(a)^{-1/10} \hat{\mathfrak{J}}^{-3/5} \psi(a)\, \kappa^{-1/10}$ encapsulates the factors depending on magnetic and transport radial profiles ($\psi(r)$ is defined by $h_{RFP}=0$). Due to (6), the limit $\xi \to 0$ of (14) is finite.

As far as the RFP is concerned, the RFX-mod scaling, $\tau_{E,RFX-\text{mod}}(ms) \approx 0.0046 \times I_p(MA)^{-0.15} \bar{n}_e(10^{19} m^{-3})^{0.37} \tilde{b}_{norm}^{-1.1}$ [27], is applied in (13). Here, $\tilde{b}_{norm}$ (of the order of $10^{-2}$) represents the normalized amplitude of the MHD perturbations, excluding the dominant mode. Then, $(\kappa \times K_*/a^2)^{1/10}$ turns out to be nearly constant over the RFX-mod database, having fluctuations of order 3% of the average. Therefore, factor $K_*^{1/10}$ is replaced by this average times $\kappa^{-1/10} a^{1/5}$ in (11):

15) $\quad n^*_{RFP} \leq 0.38 \times R_0^{1/5} Z_{eff}(0)^{2/5} f_*(\%)^{-1/2} \tilde{R}^{-1/2} B_\phi(0)^{-1/5} \Psi_{RFP} \times n_G$

Here, $\Psi_{RFP} = q(0)^{-2/5} \hat{\mathfrak{J}}^{-3/5} \psi(a) \kappa^{-1/10}$, and $\psi(r)$ is defined by $h_{RFP}$=1, $h_{AUX}$=0. Both scalings (14), (15) have a Greenwald-like structure, modulated by order unity factors. To make them quantitatively more precise, an estimate of $\Psi_{tok}$, $\Psi_{RFP}$ is needed. This may be accomplished only by resorting to some numerical evaluation. In the ohmic tokamak case ($h_{AUX}$=0, $h_{RFP}$=0, so $\psi(a)$=1), a standard current model $J \propto (1-x^2)^{q(a)/q(0)-1}$ [16], $x=r/a$, the parametrization $\hat{K} = Exp\{(1-x^2) \ln[K(0)/K_*]\}$ and the identification $\kappa = \langle \hat{K} \rangle$, allow writing $\Psi_{tok} \approx \Psi_{tok}[q(a)/q(0), K(0)/K_*]$. Magnetic and thermal profiles are not independent here: exploiting parallel Ohm's law $E_\phi B_\phi = \eta \sigma B^2 / \mu_0$, on-axis relation (6) and approximation $B^2 \approx B_\phi^2$ (tokamak ordering) one gets $q(a)/q(0) \cong 0.5 \times \left\{ \int_0^1 [T/T(0)]^{3/2} x \, dx \right\}^{-1}$. The dependence of $T/T(0) = \hat{T}$ on $K(0)/K_*$ is then studied solving (1) as an homogeneous eigenvalue equation $d(x \hat{K} d\hat{T}/dx)/dx + x \Omega \hat{T}^{3/2} = 0$, given that $\wp_{Ohm} \approx E_\phi^2 / \eta$. The radiation loss term is neglected here (it acts only in a narrow edge layer and consequently it hardly impacts on the global $\hat{T}$ profile). The eigenvalue $\Omega$ is determined by the edge condition $\hat{T}(1) = 0$. Eventually, it is found $q(a)/q(0) \approx 3.46 \times [K(0)/K_*]^{-0.3}$ for $0.1 \leq K(0)/K_* \leq 10$. Accordingly, $\Psi_{tok}$ is well fitted by a weak function of the $K$ profile: $\Psi_{tok} \approx 1.9 \times \Phi[K(0)/K_*]$, where $\Phi(x) = x^{0.3}$, for $0.1 \leq x \leq 1$, $\Phi(x) = [\ln(x)+1]^{0.28}$, for $1 \leq x \leq 10$. Taking $K(0)/K_* \approx 1$, hence $\Psi_{tok} \approx 1.9$, is a sensible approximation, since in tokamaks heat diffusivity increases with $r$ [28], whereas particle density follows the opposite trend. In the auxiliary heating case ($h_{AUX}$=1, $h_{RFP}$=0) $\Psi_{tok}$ is found having a weak dependence on the $\hat{\Gamma}$ profile, which indeed enters only within an integral relation. Since (12) entails $\hat{\Gamma} \approx 1$, even more so, we may take $\hat{\Gamma} = 1$, making $\Psi_{tok}$ the same as in the ohmic case. Finally, $\Psi_{RFP}$ is estimated for the typical magnetic and temperature profiles of RFX-mod. This requires also some lengthy data analysis whose details are given in [29]. A limited range of variation is found, thus $\Psi_{RFP}\approx2.6$.

Now, we benchmark scalings (14), (15) with experimental results. As far as ohmic tokamak is concerned, recent FTU experiments [30, 31], realized in clean machine conditions with negligible content of metals ($Z_{eff}$<1.5, D as main ion), established a Greenwald-like scaling for the edge density limit: $(n_{edge})_{DL}\approx0.35\times n_G$. The similarity with equation (14) stimulates a more quantitative comparison. The FTU geometry ($a$=0.28m, $R_0$=0.935m), and $\Psi_{tok}$=1.9, $h_{AUX}$=0, are then set in (14). A constant radial profile is taken for $Z_{eff}$ (apart the very edge where the measurement is less reliable, reconstructions in different devices both tokamak, stellarator and RFP support this assumption [32-36]), alongside the scaling $Z_{eff} \cong 1+\zeta/\bar{n}_e$ (H or D as main ion, so $Z_i$=1), which models the generally observed decreasing trend with density [34]. The constant $\zeta$ mostly depends on the wall conditions for an ohmic device. Given the experimentally observed intervals $(\bar{n}_e)_{DL}$ $\approx1\div4(10^{20}\text{m}^{-3})$, $Z_{eff}$<1.5, we fix $\zeta$=0.5($10^{20}$m$^{-3}$). The $Z_{eff}$ definition [16], combined with charge neutrality condition, allows writing $f_* \cong (Z_{eff} -1)/(Z_I^2 - Z_I)_*$, where denominator is an average impurity charge within the radiative layer. Both $(Z_I^2-Z_I)_*$ and $\tilde{R}$ are here estimated by taking an Oxygen (O), Boron (B) mixture [20], representative of clean machine conditions obtained by boronization. Equation (14) allows computing $n^*$ from the experimental $B_\phi$, $n_G$ and the above estimate of $Z_{eff}$ (made with the experimental $\bar{n}_e$). The result, weakly depending on the ratio $f_B/f_O$ among B, O concentrations, is in fairly good quantitative agreement with the experimental edge densities, as shown by figure 1.

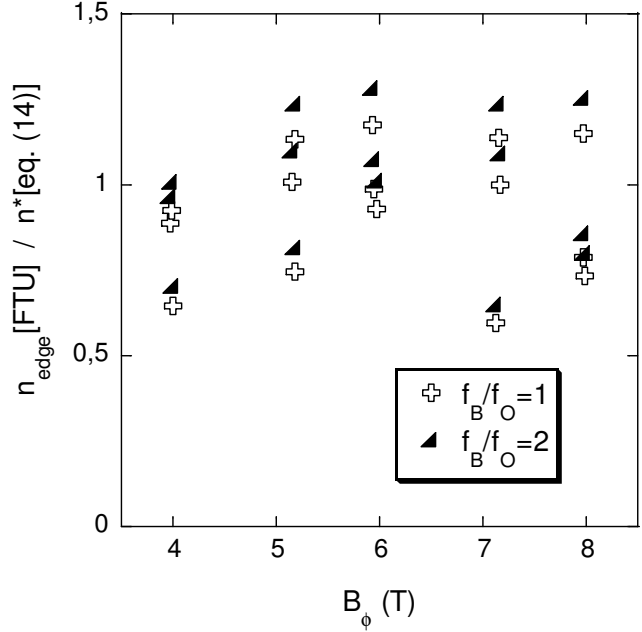

**Figure 1.** Ratio between the experimental FTU edge density at the DL disruption and the prediction of equation (14). For the same $B_\phi$ there are three different $I_p$.

As far as DL in auxiliary heated tokamak is concerned, NBI-heated L-mode discharges performed in Textor-94 support a Greenwald-scaling modified by the additional dependence $P_{tot}^{0.44}$ for both edge and line-average densities [37]. Given the nearly identical exponent for $P_{tot}$ predicted by (14) ($h_{AUX}=1$), the present model seems compatible with these results, even though a more detailed comparison is prevented by the lack of information on the current drive and $P_{Ohm}$ in [37]. The H-mode is not addressed here, since a modelling of the edge pedestal would be required, and this is beyond the scope of the present work.

RFX-mod is taken as term of comparison for the RFP. To this purpose, we need transforming (15) into a constraint on the line-average density, which is the quantity usually analysed in RFX-mod. This is accomplished by supplying the density profile peaking factor $\delta = \bar{n}_e / n^*$. The profile is rather flat, with a tendency to become hollow at high density: inversion of interferometric data [38] gives $\delta \approx 0.94 \times \bar{n}_e \left(10^{20} m^{-3}\right)^{-0.76}$. In this estimate, $n^*$ is identified by $r_*=0.95a$, which is consistent with the choice $T_*=35\mathrm{eV}$ [20] and the typical temperature profile close to the DL [29]. $Z_{eff}$ can be modelled by a constant radial profile [36], inversely scaling with density as done for the tokamak ($Z_i$=1). According to [39] one can take $\zeta \approx 0.2 \div 0.4 (10^{20} m^{-3})$. Setting the RFX-mod geometry ($a$=0.459, $R_0$=2), $\Psi_{RFP}$=2.6, and using approximation $B_\phi(0) \approx I_p(MA) = \pi a^2 \times n_G(10^{20} m^{-3})$, valid within few %, DL (15) can be cast in the following implicit form (density in $10^{20} m^{-3}$):

$$16) \quad n_G \geq 15 \times \left[ \tilde{R}/(Z_I^2 - Z_I)_* \right]^{5/8} \zeta^{5/8} \left( 1 + \frac{\zeta}{\bar{n}_e} \right)^{-1/2} \bar{n}_e^{1.575}$$

Note that the r.h.s considerably deviates from a linear function in $\bar{n}_e$. The plasma is mainly polluted by C and O, due to the presence of a graphite wall. According to [36, 40] a reasonable range of variation is $f_C/f_O \approx 2 \div 4$. In any case the dependence on $f_C/f_O$ is very weak. As term of experimental comparison for (16) we take the union of the two databases used for figures 1 of [6] and [8]: the resulting ensemble is displayed by the crosses in figure 2. The curves give and idea of the sensitivity of the model prediction to the different parameters involved. Note that the boundary of the experimental points is well delimited by equation (16).

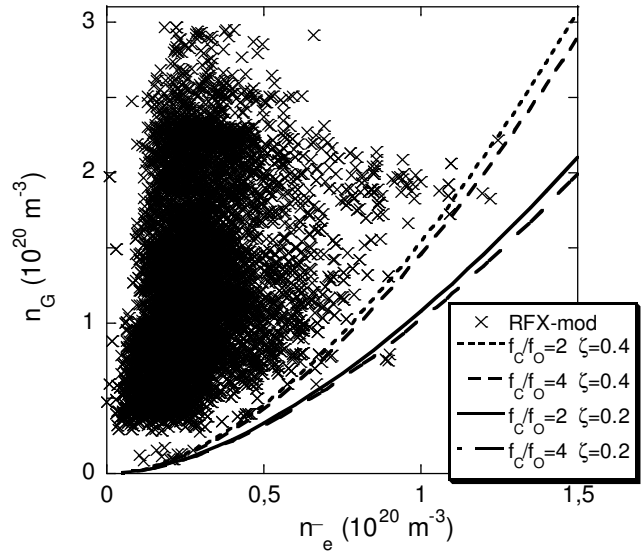

**Figure 2.** Crosses: RFX-mod data. Lines: boundary delimited by equation (16), computed for two values of both $\zeta$ and $f_C/f_O$.

*Pure auxiliary heated cylindrical configuration*

We go back to equation (5): letting $E_\phi=0$ it becomes

17) $n^* \leq \dfrac{P_{aux}}{2\pi^2 a R_0}\left(8 K_* R_* T_* f_*\right)^{-1/2}$

Unlike equation (9), $K_*^{-1/2}$ brings the total dependence on transport now, since $P_{aux}$ is mainly an external input. Therefore, the transport model takes a fundamental importance here. We use relation (13), with radially constant $K$, combined with the International Stellarator scaling 95, $\tau_E^{ISS95}(s) = 0.079 \times a^{2.21} R_0^{0.65} P_{aux}^{-0.59}(MW)\, \bar{n}_e^{0.51}(10^{19} m^{-3})\, B_\phi^{0.83}\, \iota_{2/3}^{0.4}$, [41]. With reference to LHD, we also take a radially constant $Z_{eff}$ [35], $Z_i=1$, and a mixture of C, O, with a predominance of C [42]. In terms of the line-average density, (17) gives (density in $10^{20} m^{-3}$ and power in MW):

18) $\bar{n}_e \leq 0.156 \times P_{aux}^{0.57}\, B_\phi^{0.33}\, R_0^{-0.54}\, a^{-0.72}\, \iota_{2/3}^{0.16}\, \delta^{0.80} \left(Z_{eff}-1\right)^{-0.4}$

The pre-factor dependence on $f_C/f_O$ is very weak in this case too: here we choose $f_C/f_O=3$. Experiments suggest including the peaking factor in the Sudo scaling [14], to model the fact that the actual DL refers to the edge density [15]:

19) $\bar{n}_{e,LHD} \leq 0.2 \times P_{aux}^{0.5}\, B_\phi^{0.5}\, R_0^{-0.5}\, a^{-1}\, \delta$

The similarity between (18) and (19) is evident, apart from an explicit $Z_{eff}$ dependence present in the analytical prediction. For typical LHD parameters, $R_0=3.65$m, $a=0.64$m, $B_\phi=2.71$T, $P_{aux}=2\div10$MW, $\delta=0.8\div4$ [15], $\iota_{2/3}=0.75$ [43], $1<Z_{eff}<2$ [35], the ratio between (18) and (19) varies in the range $1.3\div2$ for $Z_{eff}=1.05$, decreasing to $0.42\div0.65$ for $Z_{eff}=2$. The agreement is within a factor 2, which is rather good in the view of the cylindrical geometry approximation.

In conclusion, we have shown the existence of a DL related to the possibility of establishing a 1D magneto-thermal equilibrium in the presence of light impurity radiation losses. Though this work cannot be considered the last word on a complex and multifaceted problem such as the DL in fusion plasmas, our analysis has some strong undeniable points: 1) it is based on a careful analysis of few basic physical elements; 2) it is quantitatively consistent with the experiments taken so far as term of comparison; 3) it is transversal to the main magnetic configurations.

**Acknowledgement.** We wish to thank L. Carraro, A. Fassina, R. Paccagnella, M. Valisa, M. Zuin for several helpful discussions and suggestions, P. Innocente for providing us the RFX-mod confinement-time data-base, M. E. Puiatti and G. Spizzo for providing us the RFX-mod density limit data-bases. Finally, we wish also to thank F. Auriemma for important analyses of the RFX-mod density profile. This project has received funding from the European Union Horizon 2020 research and innovation program under grant agreement number 633053. The view and opinions expressed herein do not necessarily reflect those of the European Commission.

## Supplemental Material

Some technical issues, omitted in the corpus of the Letter, are provided in the following paragraphs.

### Section1. Cooling rates and average charges for B, C, O

Figure SM-1 plots the cooling rates (a) from the ADAS database [1], and the average charges (b) from the database [2], as function of electron temperature, for Oxygen (O), Carbon (C) and Boron (B). They are, in fact, the most relevant light species, as far as the radiation losses are concerned. The former two are intrinsic in any machine (O from vapour, C from steel or graphite tiles), the latter is often used for wall-conditioning. The charge curves are somewhat schematized with respect to the actual ones.

The *Rad(T)* curves are used to define the quantities $T_*$ and $\tilde{R} = R_* T_* \times 10^{33}$ of the step-wise model described in the Letter. We set $T_*$=35eV as the temperature delimiting the edge radiative layer: in fact at this temperature $Rad_O$ drops by a factor 10 with respect to its principal maximum, and the principal maxima of $Rad_B$, $Rad_C$ occur at lower *T*. Nevertheless we include also the secondary maximum of *Rad(T)* (or at least a portion of it) in the quantitative definition of $\tilde{R}$. In fact, we define $\tilde{R}_j = 10^{33} \times \int_0^{T(0)} Rad_j \, dT$ for *j*=O, C, B. For the FTU tokamak, one can take *T(0)*=1keV: this provides $\tilde{R}_O = 2.35$, $\tilde{R}_C = 0.7$, $\tilde{R}_B = 0.19$, in units of Wm³keV. For RFX-mod, one can take *T(0)*=0.2keV (see analysis discussed in figure SM-2): this provides $\tilde{R}_O = 1.7$, $\tilde{R}_C = 0.59$, $\tilde{R}_B = 0.15$, in units of Wm³keV.

The charge curves are used to compute the average values $(Z_j^2 - Z_j Z_i)_*$, being $Z_i$ the main ion charge, referred to the temperature interval [0, $T_*$]: for $Z_i$=1 we get 15.6, 9, 6 for *j*=O, C, B respectively.

The simultaneous presence of the different kinds of impurities is managed by defining suitable weighted-averages, both for the cooling rate and the charge. If $f_j = n_j/n_e$ denotes the *j*[th] impurity concentration, use is made of the following relations:

SM1) $$Rad = \left[ Rad_O + \sum_{j=B,C} Rad_j \, f_j/f_O \right] \left[ 1 + \sum_{j=B,C} f_j/f_O \right]^{-1}$$

SM2) $$Z_I^2 - Z_I Z_i = \left[ Z_O^2 - Z_O Z_i + \sum_{j=B,C} (Z_j^2 - Z_j Z_i) f_j/f_O \right] \left[ 1 + \sum_{j=B,C} f_j/f_O \right]^{-1}$$

SM3) $$Z_{eff} = Z_i + (Z_I^2 - Z_I Z_i) f_I, \qquad f_I = \sum_{j=O,B,C} f_j$$

For the sake of simplicity, the radial profile of the concentration is assumed to be the same for all the impurities, so the ratios $f_j/f_k$ are fixed numbers.

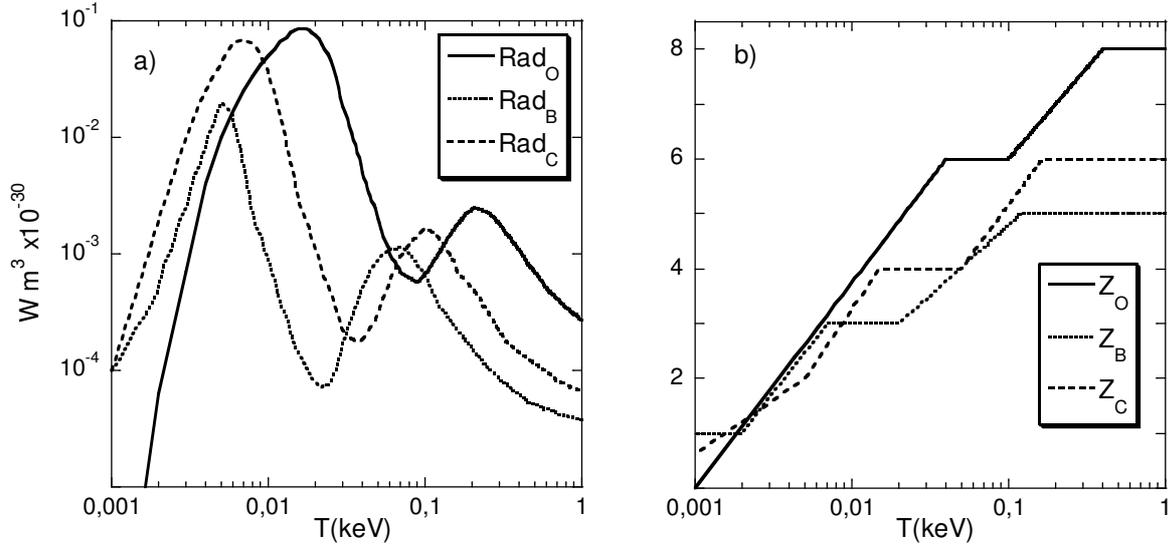

**Figure SM-1.** Plot a): cooling rates for O, B, C, as function of electron temperature, with both axes in logarithmic scale. Plot b): average charges for the same impurities, as function of electron temperature, with x-axis in logarithmic scale.

## Section 2. Estimate of $\Psi_{RFP}$.

The computation of $\Psi_{RFP} = q_0^{-2/5} \hat{\mathfrak{S}}_{RFP}^{-3/5} \kappa^{-1/10} \psi(a)$ ($\psi(r)$ defined by $h_{RFP}=1$, $h_{AUX}=0$) is here performed by adopting magnetic and heat transport models compatible with the experimental analysis of RFX-mod ($a=0.459$m, $R_0=2$m). Close to the DL, the normalized temperature $\hat{T} = T/T(0)$ is characterized by a typical profile, shown by the points of figure SM-2: a plateau, extending up to $r=|R-R_0|\approx 0.28$, is followed by a nearly constant gradient. This suggests the existence of two nearly contiguous radial regions with significantly different transport levels. Schematically we can adopt the step-wise model

SM4) $\quad \hat{K} = \begin{cases} K(0)/K_* \gg 1, & r \leq r_T \\ 1, & r > r_T \end{cases}$

being $r_T$ about the plateau radial extent. In particular, taking $r_T=0.25$ and $K(0)/K_* \geq 50$ the solutions of the normalized transport equation

SM5) $\quad d\left(x\,\hat{K}\,d\hat{T}/dx\right)/dx + x\,\Omega\,\hat{J}_\phi/\hat{C} = 0$

being $x=r/a$ and the eigenvalue $\Omega$ determined by the edge condition $\hat{T}(1)=0$, well approximate the experimental $\hat{T}$ profile (see curves of figure SM-2). Equation (SM5) results from equations (1), (7) of the letter. The radiation loss term is neglected here, since it acts only in a narrow edge layer and consequently it hardly impacts on the global $\hat{T}$ profile. A transport model similar to (SM4) has already been used to interpret the RFX-mod temperature profiles [3].

The magnetic profiles, characterized by the two parameters $F = B_\phi(a)/\langle B_\phi \rangle$, $\Theta = B_\theta(a)/\langle B_\phi \rangle$, are modelled through the function $\sigma = \sigma(0)(1-x^\alpha)$, which is standard for RFPs [4]. The normalized anomaly function is given by $\hat{C} = \hat{B}_\phi/(\hat{\eta}\,\hat{\sigma}\,\hat{B}^2)$, being $\hat{\sigma} = \sigma/\sigma(0)$. Moreover, $\hat{\eta} = \hat{T}^{-3/2}$, by taking a radially constant $Z_{eff}$. The typical $\hat{C}$ profile so obtained is shown in figure SM-3. Finally, an estimate of the shape factor $\kappa$, defined by equation (13) of the letter, is provided. It is natural writing $\tau_E$ as the sum of the energy confinement times related to the two different transport regions:

SM6) $\tau_E \approx 0.5 \times 1.6 \times 10^{-16} \bar{n}_e \left( \dfrac{r_T^2}{K(0)} + \dfrac{a^2 - r_T^2}{K_*} \right)$

For consistency with (13) it must be $\kappa \approx \left(x_T^2/\hat{K}(0) + 1 - x_T^2\right)^{-1} \cong \left(1 - x_T^2\right)^{-1}$, $x_T = r_T/a$. We also mention that, by defining the edge electron thermal diffusivity as $\chi_* = K_*/(1.6 \times 10^{-16} \times n^*)$ (see text below equation (1) in the letter: for the RFP $\chi_i \approx (m_e/m_i)^{1/2} \chi_e \ll \chi_e$ is a sensible assumption) and applying the RFX-mod scalings (discussed in the letter) to both $\tau_E$ and $\delta = \bar{n}_e/n^*$, equation (SM6) provides $\chi_*(m^2 s^{-1}) \approx 50 \times I_p(MA)^{-1.15}$, which is compatible with previous estimates (see figure 8 of [3]).

Figure SM-4 displays $\Psi_{RFP}$ computed for two different magnetic equilibria and for simulated $\hat{T}$ profile well-fitting the experimental average one (those obtained with $K(0)/K_* \geq 50$). There is no appreciable dependence both on the magnetic equilibrium, and on $K(0)/K_*$. Since the shallow reversal condition ($F \approx -0.05$) is more frequent in RFX-mod, we can take $\Psi_{RFP} \approx 2.6$.

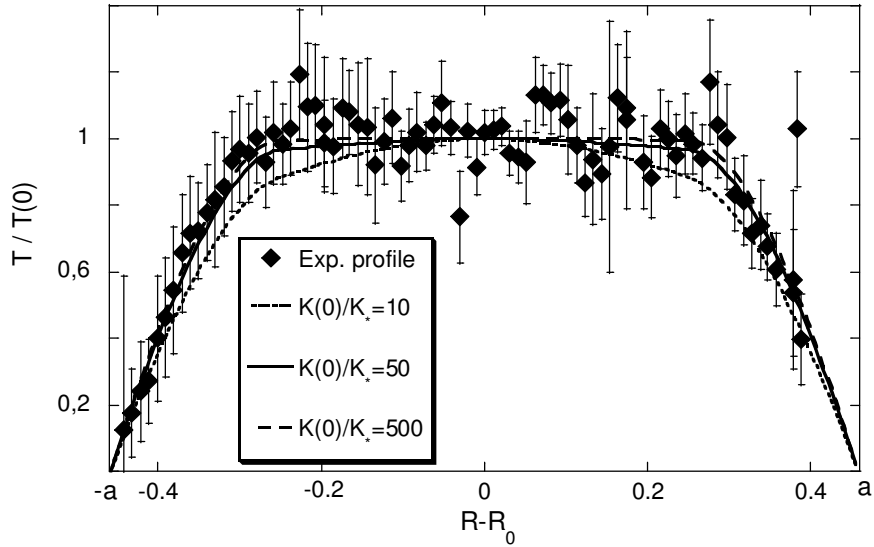

**Figure SM-2.** Diamonds: RFX-mod electron temperature radial profile measured by Thomson scattering, normalized to the average of the 4 nearest points to the magnetic axis, and then averaged over 34 shots in conditions close to the DL. The irregularities in the plateau region are likely due to spurious effect such as the detection of light from plasma-wall interaction. The x-axis is the radial coordinate of the machine system (R, Z, φ), referred to the vacuum-vessel centre. The average absolute value of the on-axis

temperature is $T(0)=174\pm72$ eV. <u>Curves</u>: simulated profiles from equation (SM5), using the step-wise model (SM4) for $\hat{K}$ with $r_T=0.25$. It is assumed $r=|R-R_0|$.

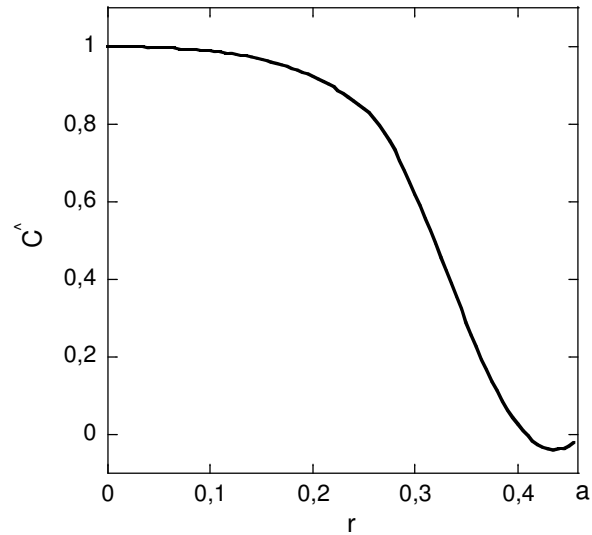

**Figure SM-3.** Estimated radial profile of the normalized anomaly function $\hat{C}$, for the equilibrium $F=-0.145$, $\Theta=1.43$.

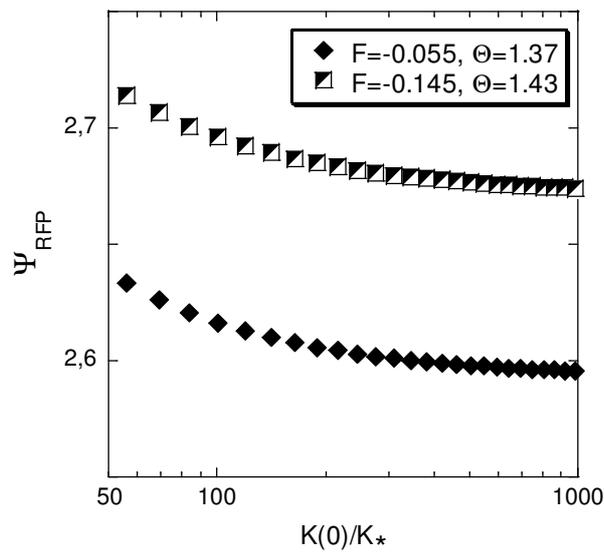

**Figure SM-4.** Estimated $\Psi_{RFP}$ as function of $K(0)/K_*$ for two different magnetic equilibria.